\documentstyle[times,pramana,epsf,floats]{ias}
\def\be{\begin{equation}}
\def\ee{\end{equation}}

\def\PL{{ Phys.\ Lett.\ }}
\def\PR{{ Phys.\ Rev.\ }}

\def\PRL{{ Phys.\ Rev.\ Lett.\ }}

\def\ZP{{ Z.\ Phys.\ }}

\newcommand{\AmS}{{\protect\the\textfont2
  A\kern-.1667em\lower.5ex\hbox{M}\kern-.125emS}}

\begin{document}
\title{Quarkonium Suppression
}
\author{
P. Petreczky}
\address{Fakult\"at f\"ur Physik, Universit\"at Bielefeld,
P.O. Box 100131, D-33501 Bielefeld, Germany}
\pacs{12.38.Mh, 11.15.Ha, 11.10.Wx, 12.39.Pn}
\keywords{quarkonium, heavy quark potential, finite temperature QCD}
\abstract{
I discuss quarkonium suppression in equilibrated strongly
interacting matter. After a brief review of basic features
of quarkonium production I discuss the application of recent
lattice data on the heavy quark potential to the problem 
of quarkonium dissociation as well as the problem of direct
lattice determination of quarkonium properties in finite
temperature lattice QCD.
}
\maketitle

\section{Introduction}
The behavior of the heavy quarkonium states in hot strongly
interacting matter was proposed as a test of its confining nature, 
since a sufficiently hot deconfined medium will dissolve  
any binding between the quark-antiquark pair \cite{matsui86}.
Another possibility of dissociation of certain quarkonium states
(subthreshold states at $T=0$) is the decay into open charm (beauty)
mesons due to in-medium modification of both the quarkonia and heavy-light
meson masses \cite{digal01a,wong}. 

The production of  $J/\psi$ and $\Upsilon$ mesons 
in hadronic reactions occurs in part
through production of higher excited $c \bar c$ (or $b \bar b$) states 
and their decay into quarkonia ground state. Since the lifetime of
different subthreshold quarkonium states is much larger than the
typical life-time of the medium which may be produced in nucleus-nucleus
collisions their decay occurs almost completely outside the produced
medium. This means that the produced medium can be probed not only by
the ground state quarkonium but also by different excited quarkonium states.
Since different quarkonium states have different sizes ( binding energies ), 
one expects that higher excited states will dissolve at lower 
temperature than the smaller and more tightly bound ground states.
These facts may lead to a sequential suppression pattern in $J/\psi$ and
$\Upsilon$ yield in nucleus-nucleus collision as the function of the
energy density \cite{seq}.

The rest of the paper is organized as follows. In section 2 I 
briefly review the basic features of quarkonium production. In section
 3 I discuss the possibility of quarkonium dissociation below
deconfinement. In section 4 I use the lattice data on the heavy quark
potential to determine the dissociation temperatures due to color
screening. Finally the problem of direct lattice determination
of quarkonium properties at finite temperature is discussed in
section 5 followed by the conclusions presented in section 6.

\section{Quarkonium production and feed-down}

It is well known that $J/\psi$ production in hadron-hadron collision is due to 
a considerable extent to the production and subsequent decay of higher
excited $c \bar c$ states \cite{cobb78,lem82,ant92}. The feed-down from
higher excited states was systematically studied in proton-nucleon and
pion-nucleon interactions with $300 GeV$ incident proton (pion) beams 
\cite{ant92}. In these studies the cross sections for
direct production of different charmonium states (excluding feed-down)
were measured. Then making use of the known branching ratios 
$B[\chi_1(1P) \to \psi(1S)] = 0.27
\pm 0.02$, $B[\chi_2(1P) \to \psi(1S)] = 0.14 \pm .01$, and
$B[\psi(2S) \to \psi(1S)] = 0.55 \pm 0.05$, one obtains the fractional
feed-down contributions $f_i$ of the different charmonium states to the
observed $J/\psi$ production; these are shown in the second and third columns of
Tab. 1

\begin{center}
\begin{tabular}{|c|c|c||c|c|c|}
\hline
      &          &              &              &          &     \\
state & $f_i(\pi^-N)$ [\%] & $f_i(p~N)$ [\%]& state & $f_i(p\bar p)$ [\%]
& $f_i^{NRQCD}(p \bar p)$ [\%] \\
&  &  & & &  \\
\hline
\hline
&  &  & & &  \\
$J/\psi(1S)$ & 57 $\pm$ 3 & 62 $\pm$ 4  & $\Upsilon(1S)$
& 52 $\pm$ 9 & 52 $\pm$ 34  \\
&  &  & & & \\
\hline
&  &  & & &  \\
$\chi_1(1P)$  & 20 $\pm$ 5  & 16 $\pm$ 4  & $\chi_b(1P)$ &
26 $\pm$ 7 & 24 $\pm$ 8  \\
&  &  & & &  \\
\hline
&  &  & & &  \\
$\chi_2(1P)$ & 15 $\pm$ 4 & 14 $\pm$ 4 & $\Upsilon(2S)$ & 10
$\pm$ 3
& 8 $\pm$ 7  \\
& &  & & & \\
\hline
& &  & & & \\
$\psi(2S)$ & 8 $\pm$ 2 & 8 $\pm$ 2 & $\chi_b(2P)$ & ~10 $\pm$ 7
&
14 $\pm$ 4 \\
& &  & & & \\
\hline
\hline
& &  & & & \\
 &  &  & $\Upsilon(3S)$
 & 2 $\pm$ 0.5  & 2 $\pm$ 2\\
& &  & & & \\
\hline
\end{tabular}
\bigskip
\par
\centerline{Table 1: Feed-down fractions for from higher excited states to
the $J/\psi$ and $\Upsilon$ states.}
\bigskip
\end{center}

In the case of bottomonium the experiment provides only the inclusive
(i.e. including also the feed-down from higher states) cross section 
for different $(nS)$ states \cite{cdf}. The feed-down from $(nP)$ states is known
only for transverse momenta $p_T \ge 8 GeV/c$ \cite{cdf}. To analyze the complete
feed-down pattern, we thus have to find a way to extrapolate these data
to $p_T=0$ as well as to determine the direct cross section for
different $(nS)$ states. This can be done using the most simple
and general model for quarkonium production, the color evaporation model \cite{mangano}.
In particular this model predicts that the ratios of cross sections for production
of different quarkonium states are energy independent. This prediction
was verified for a considerable range of energies \cite{gavai95}. 
The ratios between the different $\chi_J(1P)$ states in this model are predicted to be
governed essentially by the orbital angular momentum degeneracy
\cite{mangano}; we thus expect for the corresponding cross-sections
\be
\chi_0(1P) : \chi_1(1P) : \chi_2(1P) = 1 : 3 : 5.
\label{2.3}
\ee
From Table 1 we have for $\pi^-N$ collisions $\chi_2(1P) / \chi_1(1P)
\simeq 1.44 \pm 0.38$ and thus reasonable agreement with the predicted
ratio 1.67. Actually, for $pN$ interactions, the experiment measures only the
combined effect of $\chi_1$ and $\chi_2$ decay (30 \% of the overall
$J/\psi$ production); the listed values in Tab.~1 are obtained by distributing this in
the ratio 3:5.

Using considerations based on color evaporation model, in particular
Eq. (1), the feed-down from higher excited $b \bar b$ states to $\Upsilon$
production can be predicted \cite{digal01b}; the feed-down fractions are summarized 
in Tab.1. Alternatively the feed-down fraction from higher excited $b \bar b$
states can be predicted using NRQCD factorization formula \cite{digal01b}.
The results of this analysis are summarized in the last column of Tab. 1

\section{Quarkonium dissociation below  deconfinement}
Recent lattice calculations of the heavy quark potential 
show evidence for the string breaking at finite temperature \cite{karsch01}.
On the lattice the potential is calculated from the Polyakov loop correlator,
to which it is related by 
\be
V(r,T)=-\ln< L(r) L^{\dagger}(0)>+C,
\label{polcorr}
\ee
where $L(r)$ is the Polyakov loop (see e.g. Ref. \cite{karsch01}
for definition). 
The normalization constant $C$ contains both the cut-off dependent self-energy
and the entropy contributions $-TS$ (for $T \ne 0$, 
$-\ln< L(r) L^{\dagger}(0)>$ is
actually the free energy of the static $Q \bar Q$ pair).
For a properly chosen
normalization constant $C$, $V(r,T)$ is the ground state energy of 
an infinitely heavy $Q \bar Q$ pair.
In absence of dynamical quarks (quenched QCD) $V(r,T)$ is linearly rising with
$r$ for large separations indicating the existence of a flux tube (string).
If dynamical quarks are present the flux tube can decay (the string can break)
by creating a pair
of light quarks $q \bar q$ from the vacuum once $V(r,T)$ is larger then twice
the binding energy of a heavy-light $Q \bar q$ ($q \bar Q$) meson
\footnote{Similar phenomenon occurs of course at $T=0$. However it is 
much more difficult to observe it on lattice (see e.g. \cite{bali00}).}.
Thus the potential at very large distances is constant $V_{\infty}(T)$ and is
equal to twice the binding energy of a heavy-light meson. 

\begin{figure}
\vspace*{-0.5cm}
\epsfxsize=8cm
\centerline{\epsffile{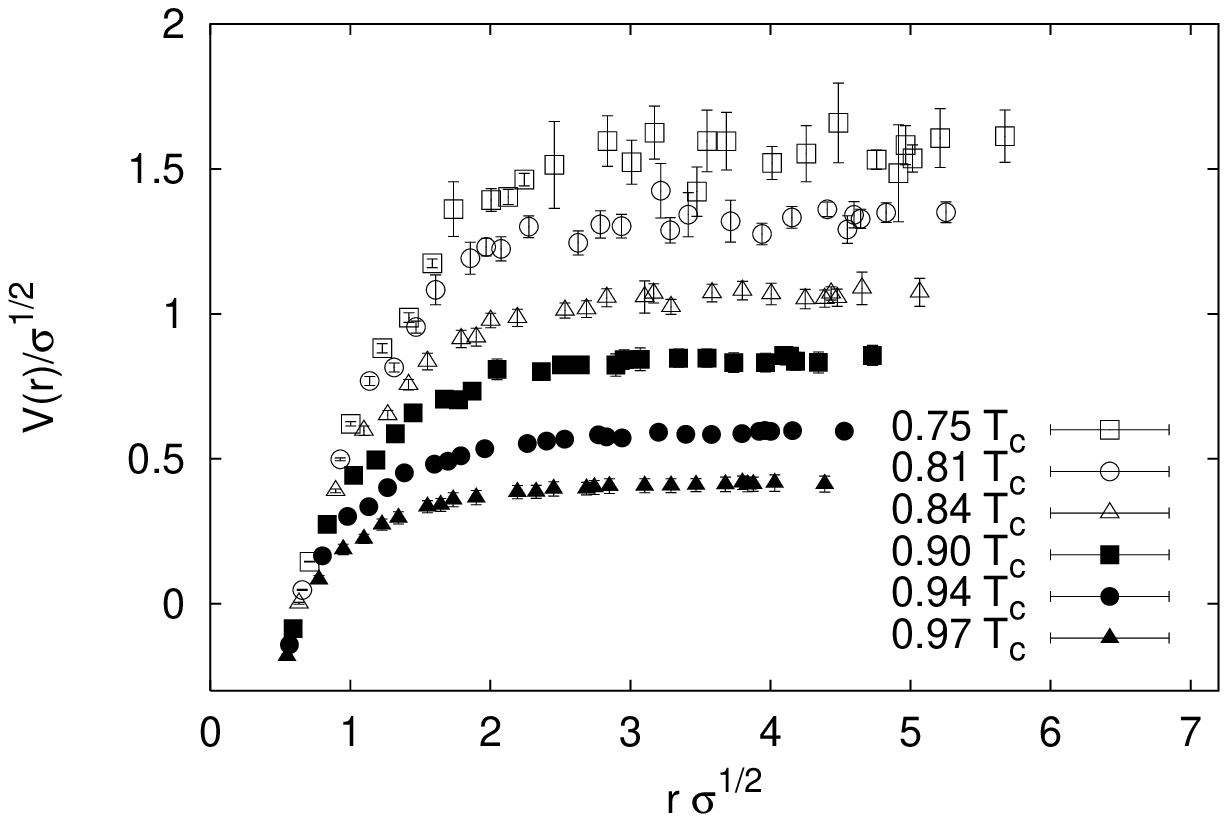}} 
\epsfxsize=8cm 
\centerline{\epsffile{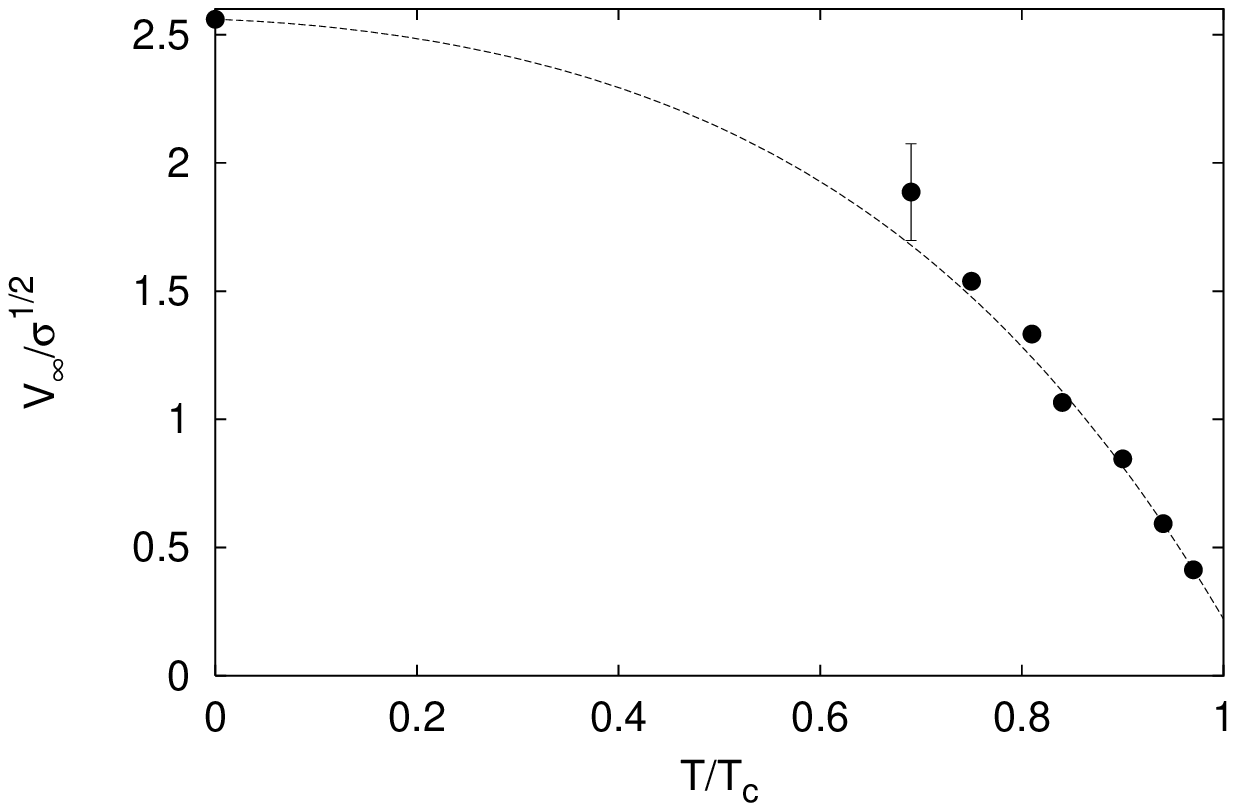}}
\vspace*{-0.3cm}
\caption{The heavy quark potential and its asymptotic value below deconfinement at
different temperatures. The line on the bottom figure is a fit to the data
points.}
\label{pot_below}
\end{figure}

It is expected that medium effects are not important at very
short distances. Therefore at very short distances the potential
$V(r,T)$ should be given by the Cornell potential \cite{eichten80}
\be
V(r)=-\frac{e}{r}+\sigma r 
\label{cornell}
\ee
We use this fact to determine the normalization constant $C$ and set the 
potential to be of the Cornell form at the the smallest distance $r T=0.25$ available
in lattice studies of \cite{karsch01},
with $e=0.4$ as expected for (2+1)-flavor QCD \cite{bali97}.
The resulting potential and $V_{\infty}(T)$ are shown in Fig. \ref{pot_below}.
Note the strong temperature dependence of $V_{\infty}(T)$.
Since for
sufficiently heavy quarks ($m_Q \gg \Lambda_{QCD}$) it does not matter whether the quark
is infinitely heavy or just merely heavy, the open charm (beauty) meson masses are
approximately given by $2 M_{D,B}(T)=2 m_{c,b}+V_{\infty}(T)$. 

\begin{figure}
\vspace*{-0.5cm}
\epsfxsize=8cm
\centerline{\epsffile{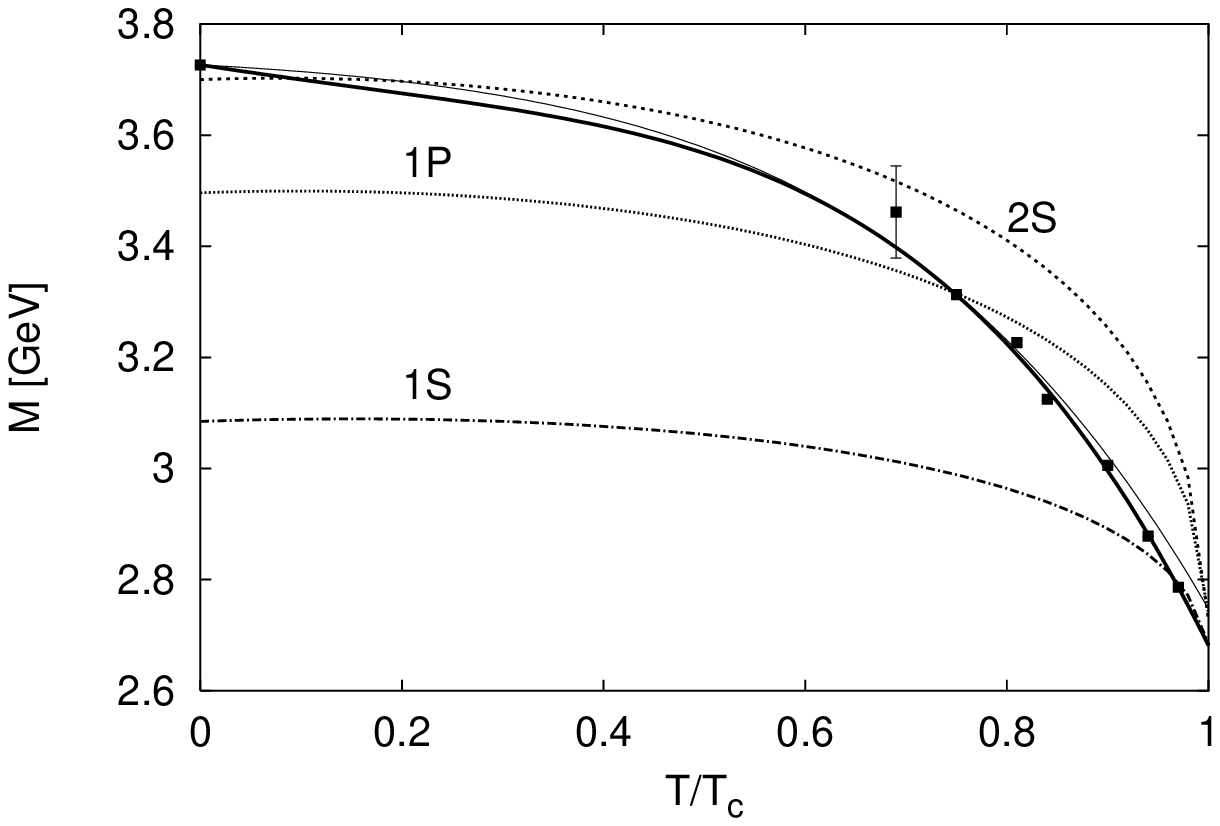}}
\epsfxsize=8cm 
\centerline{\epsffile{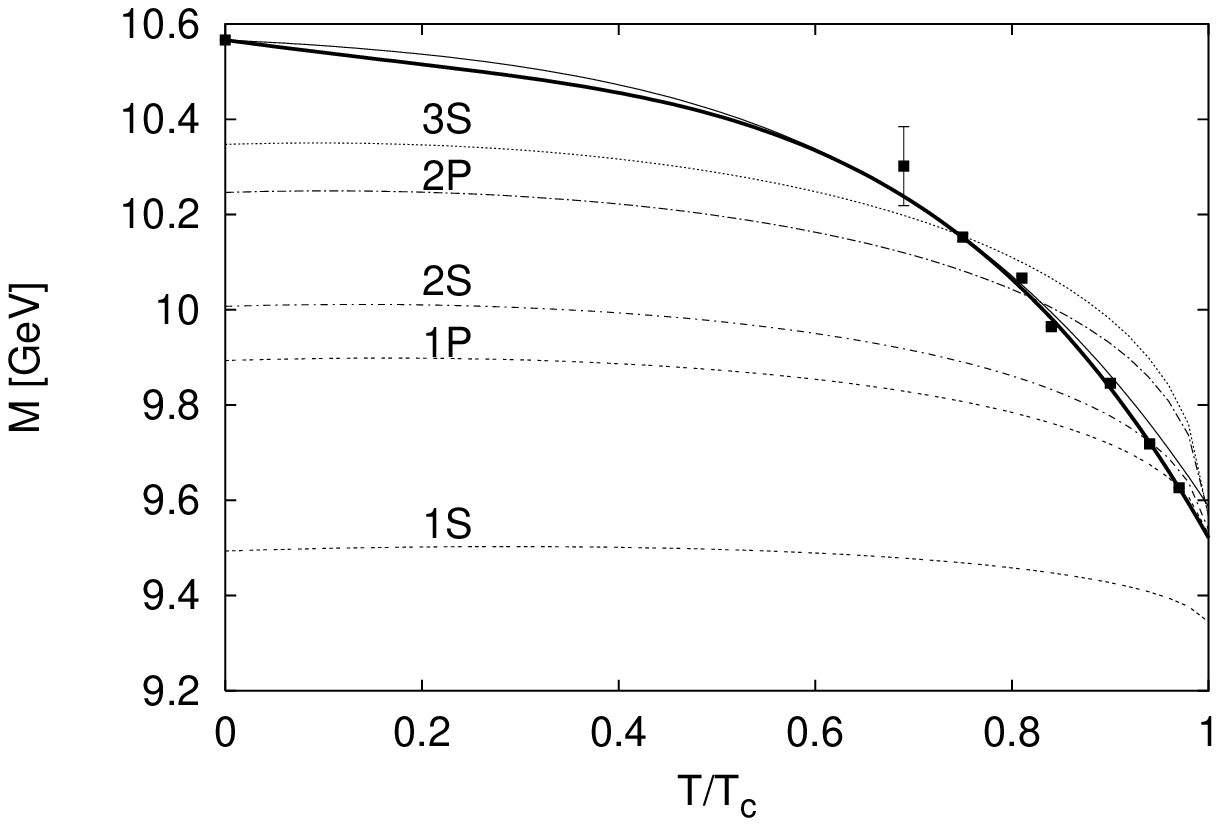}}
\vspace*{-0.3cm}
\caption{The masses of different quarkonia states and the open charm (beauty)
threshold as function of the temperature. Shown are the charmonia masses and 
open charm threshold (top) and bottomonia masses and open beauty threshold
(bottom) as function of the temperature. The thick solid line is the open charm
(beauty) threshold obtained from normalization at $r=1/(4T)$. The thin solid line
in the open charm (beauty) threshold obtained from normalization at $r=\sqrt{2}/(4T)$
(see text).}
\label{th}
\end{figure}

Now the temperature dependence of the different quarkonium states should
be addressed. At zero temperature the heavy quark masses permit the
application of potential theory for the description of quarkonium spectroscopy
(see e.g. \cite{bali00}). Furthermore it turns out that the time scale
of gluodynamics relevant for quarkonium spectroscopy is smaller
than ${(m_Q v)}^{-1}$ ($v$ being the heavy quark velocity) \cite{bali00}. 
For sufficiently
heavy quarks this time scale is much smaller than the typical hadronic
time scale $\Lambda_{QCD}^{-1} \sim 1 fm$. The decay of the flux-tube
like all other hadronic decays has time scale of order $1 fm$. Therefore
in the potential theory the potential must always  be of Cornell form 
(i.e. linearly rising at large distances). These considerations have direct
phenomenological support. Namely, simple potential models with linearly 
rising potential can describe reasonably well also the quarkonium states 
above the open charm (beauty) threshold. Many of these higher excited
states have an effective radius of order of 
or even larger than $1fm$ \cite{bali97,eichten80}.
Contrary to this situation in the case of the potential becoming flat
around $1 fm$ (the expected radius of string breaking at $T=0$) the higher excited states
above the open charm (beauty) threshold simply do not exist.
Therefore the temperature dependent quarkonium masses have been determined
from the Schr\"odinger equation
\be
\left[ 2m_Q - {1\over m_Q}\nabla^2 + V(r) \right] \Phi_i = M_i
\Phi_i,
\label{schroe}
\ee 
where the potential $V(r)$ is identified 
with the temperature dependent string potential
\cite{gao}
\begin{eqnarray}
V_{string}(r,T)&=&-(e-{1\over 6} {\rm arctan}(2 r T) )~{1\over r}+
\nonumber\\
&&
(\sigma(T)-{\pi T^2\over 3}-{2 T^2\over 3} {\rm arctan}{1\over r T})~r+
{1\over 2} \ln(1+4 r^2 T^2).
\label{vTdep}
\end{eqnarray}
This form of the potential describes quite well the temperature
dependence of the heavy quark potential in quenched QCD  for appropriately
chosen $\sigma(T)$ \cite{kaczmarek00}. In order to make contact to real QCD we set $e=0.4$
and use $T_c/\sqrt{\sigma}=0.425$ from \cite{karsch01}
for the deconfinement temperature ( by $\sigma$ we always
denote the string tension at zero temperature). 
Furthermore we use the following values of the heavy quark masses, $m_c=1.3 GeV$
and $m_b=4.72 GeV$ as well as $\sqrt{\sigma}=0.44GeV$ for the zero temperature
string tension. 
This set of parameters gives a fairly good description of the observed quarkonium
spectrum at zero temperature.
The temperature dependence of the string tension was taken from
\cite{kaczmarek00}. The resulting quarkonia masses are shown in Fig \ref{th}.

Since the smallest distance available
on lattice is only $0.25 T^{-1}$ one may worry about possible medium effects at this
distance and their role in determination of $V_{\infty}(T)$. 
Normalizing the Polyakov loop correlator (\ref{polcorr}) at $r=1/(4T)$ with
Eq.\ (\ref{vTdep}), we thus obtain what might be a more reliable estimate
of the plateau $V_{\infty}(T)$ than with the $T=0$ form (\ref{cornell}).
It turns out, however,
that the two forms of
short distance behavior resulting from the zero temperature Cornell
potential  (\ref{cornell}) and (\ref{vTdep})
are practically identical, so that the normalization is in fact not affected
by the in-medium modifications at larger distances.
To consider further possible uncertainties of the normalization procedure,
the Polyakov loop correlator has been normalized also at the next smallest
distance $r=\sqrt{2}/(4T)$. The resulting two forms of $V_{\infty}(T)$
are shown in both Fig. \ref{th}. The difference
between the two curves of $V_{\infty}(T)$ provides an estimate of
the normalization error. Except for the region very near $T=T_c$,
the uncertainty is seen to be quite small.

From Fig. \ref{th} one can see that $\psi'$ and $\chi_c$ states 
become an open charm states well below $T_c$ and can dissociate
by decaying into $D \bar D$. The situation is similar for $\Upsilon(3S)$
and $\chi_b(2P)$ states which can decay into $B \bar B$
below $T_c$. For $J/\psi$, $\chi_b(1P)$ and $\Upsilon(2S)$
it is not possible to say whether they will dissociate above $T_c$ 
or just below $T_c$. Finally, the $\Upsilon(1S)$ state will definitely
dissociate above the deconfinement. We are going to estimate the dissociation
temperatures of these states in the next section.

\section{Quarkonium dissociation by color screening and
the sequential suppression pattern}
\begin{figure}
\vspace*{-0.5cm}
\epsfxsize=8cm
{\epsffile{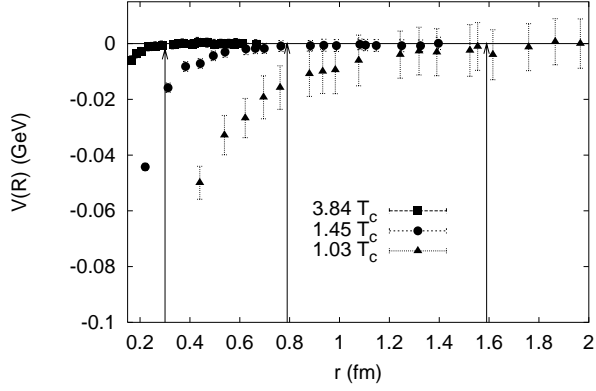}}
\vspace*{-0.3cm}
\caption{The color average potential calculated on lattice 
}
\label{above}
\end{figure}

In the deconfined phase it is customary to choose the constant $C$
in (\ref{polcorr}) to be the value of the correlator at infinite separation
$C={T \ln<L(r) L^{\dagger}(0)>|}_{r \rightarrow \infty} \equiv \ln {|<L>|}^2$.
The resulting connected correlator defines the so-called color
averaged potential \cite{nadkarni86}
\be
V(r,T)=-T \ln \frac{<L(r) L^{\dagger}(0)>}{{|<L>|}^2}
\label{vav}
\ee
The color averaged potential can be written as the thermal 
average of the potentials in color singlet $V_1(T,r)$ 
and color octet $V_8(T,r)$ states:
\be
\exp(-V(r,T)/T)=\frac{1}{9}\exp(-V_1(r,T)/T)+\frac{8}{9}\exp(-V_8(r,T)/T)
\label{av}
\ee
In potential models it is assumed that quarkonium is dominantly a singlet 
$Q\bar Q$ state. Furthermore the octet channel is repulsive (at least in
perturbation theory) and therefore only a singlet $Q\bar Q$  pair can be bound
in the deconfined phase. Thus we need to know the singlet potential.
Lattice data in the relevant case of 3 flavor QCD exist only for the
averaged potential \cite{karsch01,private}. 
The averaged potential in 3 flavor QCD 
is shown in Fig. \ref{above} for three representative temperatures.
Note that within the present accuracy of the lattice calculations the potential
vanishes beyond some distance $r_0(T)$ denoted by vertical  arrows in Fig. \ref{above}.
In perturbation theory, the leading terms
for both singlet and octet potentials are of Coulomb form 
at high temperature and small $r$ ($r << T^{-1}$)
\be
V_1(r,T) = -{4\over 3} {\alpha(T) \over r}, ~~~
V_8(r,T) = +{1\over 6} {\alpha(T) \over r},
\label{3.9}
\ee
with $\alpha(T)$ for the temperature-dependent running coupling.
In the region just above the deconfinement point $T=T_c$, there will
certainly be significant non-perturbative effects of unknown form.
We attempt
to parameterize the existing non-perturbative effects by
the following form
\be
V_1(r,T) = -{4\over 3} {\alpha(T) \over r} \exp\{-\mu(T)r\},~
V_8(r,T) = {1\over 6} c(T) {\alpha(T) \over r} \exp\{-\mu(T)r\}
\label{3.10}
\ee
where $\mu(T)$ denotes the effective screening mass in the deconfined
medium.
We fit the lattice data of \cite{karsch01,private} 
by Eqs. (\ref{av}) and (\ref{3.10}) assuming
$\alpha(T)$ and $\mu(T)$ and $c(T)$ to be unknown functions of $T$.
Since the present data are not precise enough to determine
$\alpha(T)$ and $\mu(T)$ and $c(T)$ simultaneously, some additional
constraints coming from simulations of pure gauge 
theory \cite{heller95,heller98,cucchieri01,attig88} 
should be invoked in the fit procedure. The fit procedure is 
described in detail in Ref. \cite{digal01b}. Here it is sufficient to 
mention that $\alpha(T)$ can be well described by 1-loop 
formula for the QCD running 
coupling with $\Lambda_{QCD}=(0.34 \pm 0.01)T_c$ and
the screening mass $\mu(T)$ is constant in units of the temperature
$\mu(T) = (1.15 \pm 0.02)  T$ \cite{digal01b}.

Now we are in a position to discuss quarkonium dissociation due
to color screening. It is natural to assume that the heavy $Q \bar Q$
pair cannot exist as a bound state if its effective binding radius (
the mean distance between $Q$ and $\bar Q$) is larger than the screening
radius of the medium. The effective radii for different bound states are
calculated from Eq. (\ref{schroe}) with
$V(r)=V_1(r,T)$. 
\begin{figure}
\vspace*{-0.5cm}
\epsfxsize=8cm
{\epsffile{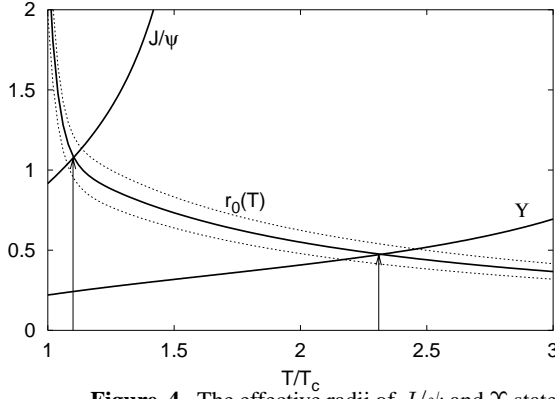}}
\vspace*{-0.3cm}
\caption{The effective radii of $J/\psi$ and $\Upsilon$ states and
the screening radius $r_0(T)$. The
dotted lines denote the uncertainty in the value of $r_0(T)$. 
}
\label{radii}
\end{figure}
The screening radius of the medium can be identified with $1/\mu(T)$.
However, the value of $\mu(T)$ strongly depends on assumptions we have
made to determine it. A less model dependent and more conservative
approach would be to identify the screening radius with $r_0(T)$
defined above. We use the latter approach. 
In Fig. \ref{radii} we show the effective radius of $J/\psi$ and 
$\Upsilon$ states and the screening radius as function of the 
temperature. The intersection of these curves defines the dissociation
temperature of $J/\psi$ and $\Upsilon$ states. Similar analysis was done
for excited states which may survive above $T_c$. 

Alternatively
one can define the dissociation temperatures as the temperature where
the effective bound state radius diverges \cite{karsch88} and the
quark-antiquark pair is unbound. It is clear from Fig. \ref{radii}
such definition will lead to larger, though not very different value
of the dissociation temperature. However, one should note that
once the radius of the bound state is larger than the screening radius
the present treatment based Schr\"odinger equation is clearly not valid
and the effect of the medium becomes so strong that quarkonium
dissociation is very likely to happen.

\begin{figure}
\vspace*{-0.3cm}
\epsfxsize=6.0cm
\centerline{\epsffile{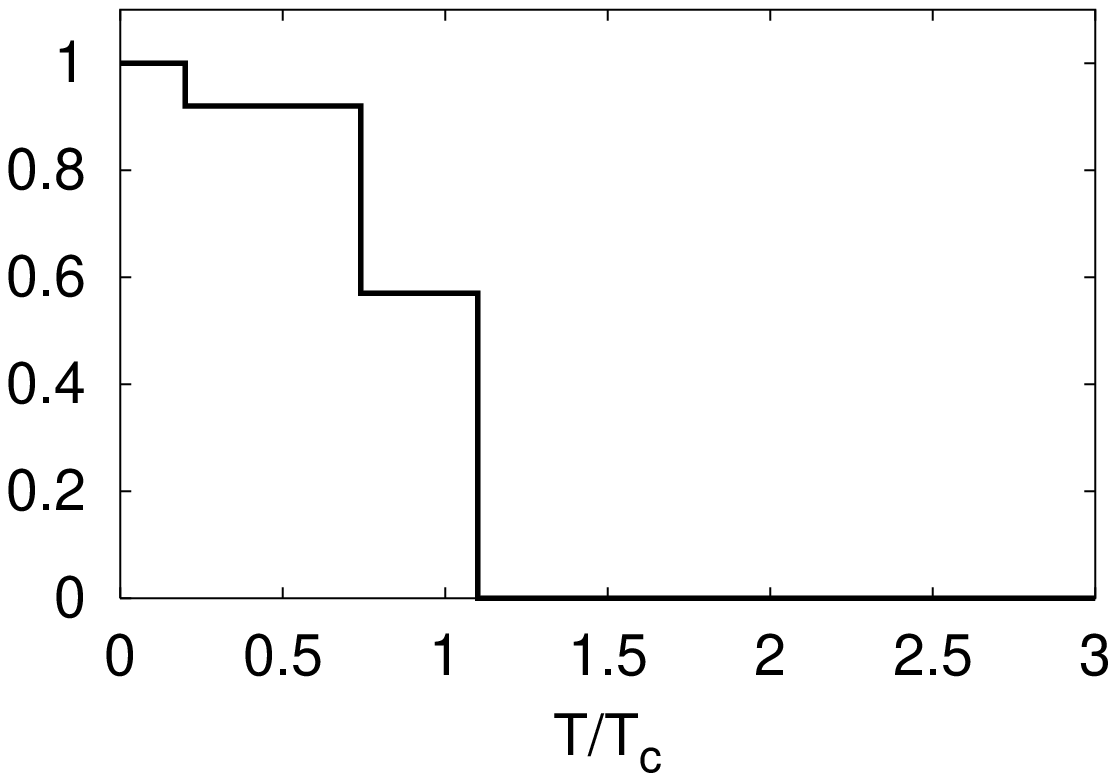} \hspace*{0.3cm}
\epsfxsize=6.0cm
\epsffile{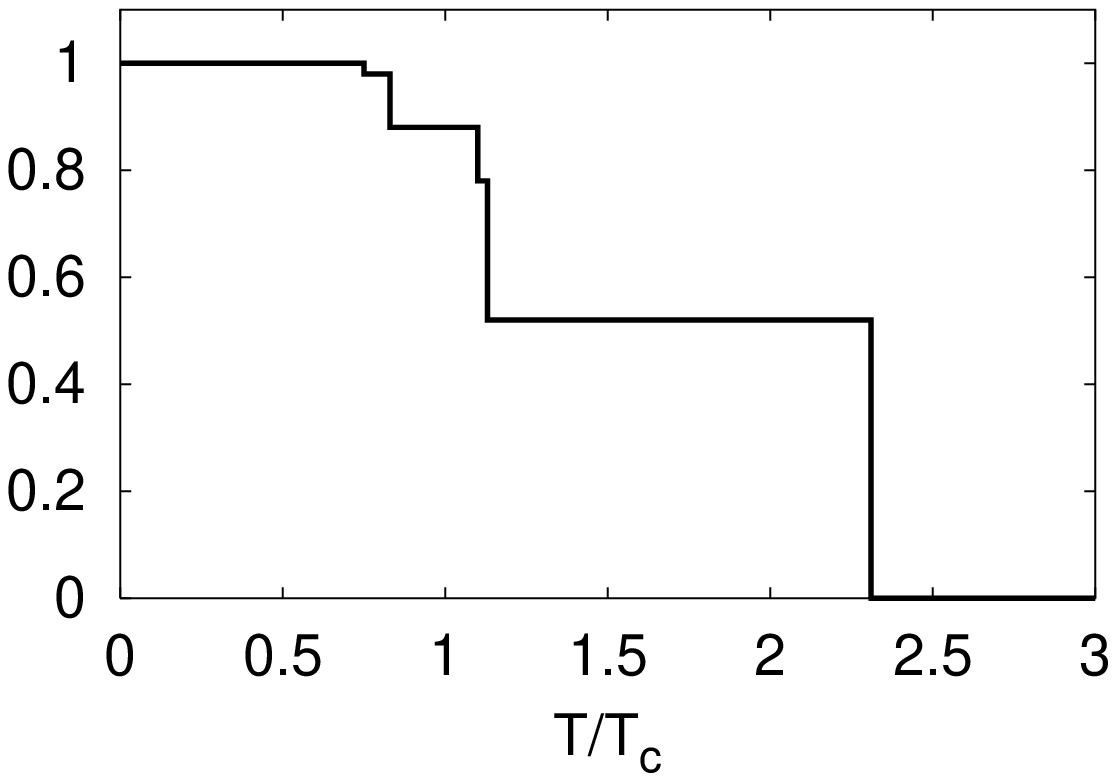}}
\caption{
The suppression pattern of $J/\psi$ (left) and $\Upsilon$ yield (right)
as function of the temperature.
}
\label{supp}
\end{figure}

Now the dissociation temperatures are known for all quarkonium states 
and  
summarized in Tab. 2. 
Combining
these dissociation temperature with the feed-down 
fractions determined in section 2
we can predict the sequential suppression pattern of $J/\psi$ 
and $\Upsilon$ states
as function of the temperature.
These are summarized in Fig. \ref{supp}.

\vspace*{0.2cm}
\begin{center}
\begin{tabular}{|c|c|c|c|c|c|c|c|c|}
\hline
$q\bar{q}$&$J/\Psi$&$\chi_c $&$\psi'$&$\Upsilon(1S)$&$\chi_b
(1P)$&$\Upsilon (2S)$&$\chi_b (2P)$&$\Upsilon (3S)$ \\
\hline
$T/T_c$&1.10 & 0.74 & 0.2& 2.31& 1.13& 1.10& 0.83& 0.75\\
\hline
\end{tabular}
\vspace*{0.4cm}

Table 2: Dissociation temperatures of different quarkonium states.
\end{center}

\section{Lattice determination of quarkonium properties
at finite temperature}

At zero temperature quarkonium properties (masses , decay constants
etc.) can be obtained from the behavior of the mesonic correlators
at large Euclidean time separations
\be
< O(\tau,\vec{p}), O(0, -\vec{p}) >_{\tau \rightarrow \infty}
= \sum_n A_n e^{-m_n \tau}
\label{t0}
\ee
here $O(\tau,\vec{p})=\sum_{\vec{x}} e^{i \vec{p} \vec{x}} O(\tau,x)$
is the mesonic operator bilinear
in quark fields $O(\tau,\vec{x})=\bar q(x) \Gamma q(x)$;
$\Gamma=\gamma_5, \gamma_{\mu}$ for the pseudo-scalar and the
vector channel,
respectively. Direct application of Eq. (\ref{t0}) for
determination of the quarkonium masses is difficult because at large
separations statistical errors in the correlation function become
very large. To improve the situation one usually uses smeared (improved)
operators $\tilde O$ instead of $O$ such that only one 
coefficient $A_n$  in Eq. (\ref{t0}) is significantly different from 
zero. A possible choice of the operator $\tilde O$ which has an optimal
projection onto a given quarkonium state is
\be
\tilde O(\tau,\vec{z})=\sum_{\vec{y}} \omega(\vec{y}) \bar q(\tau,\vec{z})
\Gamma q(\tau,\vec{z}+\vec{y}),
\ee
where $\omega(\vec{y})$ is the trial wave function
\footnote{This definition works only in Coulomb gauge} (see
\cite{bernard} for further details). 

At finite temperature further  difficulties appear in direct
lattice determination of quarkonium masses. First of all
Eq. (\ref{t0}) is not applicable in principle because the time
extent is limited by the inverse temperature. Since the temporal lattice
size becomes smaller as the temperature is increased 
fewer data points on the temporal correlators are available. 
Another problem which is also present at zero temperature but becomes
more serious at finite temperature (just because one is enforced to
consider mesonic correlators also at short distances) is the
discretization errors of order $m_Q a$ with $m_Q$ being the
heavy quark mass and $a$ being the lattice spacing.

To study quarkonium properties at finite temperature 
in quenched QCD
correlators of smeared operators were calculated on anisotropic
lattices \cite{fingberg,umeda}. To deal with large
discretization errors in Ref. \cite{fingberg} the NRQCD 
formalism was used (where the scale $m_Q$ is integrated out).
In this calculation the mesonic correlators corresponding to
the ground state quarkonium show only small changes up to
temperature $1.2 T_c$ compared to the zero temperature case
while the  first excited state shows very dramatic change with
the temperature. The authors of Ref. \cite{umeda} used 
a different formalism, namely they used the Fermilab action
\cite{elk} on anisotropic lattices which has no discretization error
of order $m_Q a$ at tree level. They observed more dramatic
changes of the mesonic correlators as the deconfinement point is
crossed, however, the correlators are very far from the free ones
even at temperatures as high as $1.5 T_c$ indicating  possible
existence of bound states at this temperature. 

There are several problems with the approach presented in Ref.
\cite{fingberg,umeda}. The most serious one is the use of 
optimized correlators which assumes existence of well defined
quarkonium states at finite temperature. Clearly, different particle
states at zero temperature will appear as quasiparticles with finite
width at non-zero temperature.
The implementation of NRQCD formalism 
in \cite{fingberg} does not assume anti-periodic boundary condition 
in time direction for
the quark propagators and therefore it is not clear to what extent the
meson correlators calculated in \cite{fingberg} can be related to some
finite temperature (retarded) correlation functions. In Ref. \cite{umeda}
the spatial lattice spacing was quite large and the effect of doublers
cannot be completely neglected. 

Another possibility to extract quarkonium properties in lattice QCD is
to calculate usual point-to-point correlators 
$G(\tau,\vec{p})=< O(\tau, \vec{p}) O(0, -\vec{p}) >$ (i.e. correlators of point
sources) in imaginary
time and extract the spectral function. The information on
the bound state are then encoded in the peaks of the spectral function.

The imaginary time correlator can be related to the retarded meson
propagator $G_R(\omega,p)$ by analytic continuation
\be
G(i \omega_n,p)=\int_0^{1/T} e^{i \omega_n \tau} G(\tau,p),~~~
G(i \omega_n \rightarrow \omega +i \epsilon,p)=G_R(\omega,p).
\ee
which allows to write down the spectral representation for $G(\tau,p)$
\be
G(\tau,p)=\int_0^{\infty} d \omega 
\sigma(\omega,p) \frac{\cosh(\omega(\tau-1/(2
T))}{\sinh(\omega/(2 T))}
\ee
In principle this equation allows to determine $\sigma(\omega)$ from
$G(\tau,p)$. However, in practice the value of $G$ is available only
for $N_{\tau}/2 \sim 8$ different $\tau$ -values. For reasonably fine 
discretization in $\omega$ -space one has $N_{\omega} \sim 700$ degrees
of freedom to be reconstructed. This problem can be solved only
using the {\em Maximum Entropy Method} (MEM) (see \cite{asakawa} for review).
Mesonic spectral functions in different channels were successfully
reconstructed using this method \cite{hatsuda}. Very recently it 
has been demonstrated that spectral function also at finite temperature
can be reconstructed using this method 
and isotropic lattices with $N_{\tau}=12,~16$ sites in temporal
direction \cite{wetzorke01,karsch01a}. Thus a natural alternative to
the approach presented in Ref. \cite{fingberg,umeda} could be the
calculation of point-to-point mesonic correlators using isotropic
lattice with non-perturbatively improved clover action \cite{karsch01a}.
Compared to Ref. \cite{umeda} this approach has the advantage that
all lattice artifacts of order $Ta$ $pa$ are completely removed but
the also the disadvantage that $m_Q a$ effects are present even at the tree
level. The latter, however, can be controlled if $a$ is small enough.
This approach is currently being investigated \cite{datta}. 
Calculations so far have been performed on $48^3\times 16$, $48^3 \times 12$
and $64^3 \times 16$ lattices and gauge coupling $\beta=6.499,~6.640$ and 
$6.872$. These parameters correspond to temperatures $T=0.9T_c,~1.2 T_c$ and
$1.5T_c$ and lattice spacing from $a^{-1}=6.5GeV,~4.9GeV$ and
$4.0 GeV$. 
\begin{figure}
\epsfxsize=8cm
\centerline{\epsffile{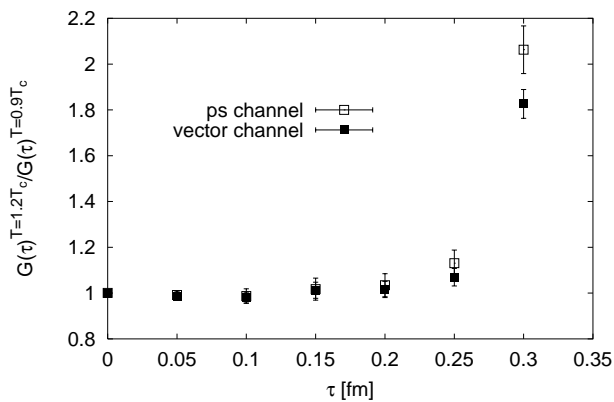}}
\caption{The ratio of the correlators in pseudo-scalar and vector channels
at $1.2 T_c$ to the one at $0.9T_c$. Calculations were done on
at $\beta=6.499$ on $48^3\times 12$ and $48^3\times 16$ correspondingly.}
\label{ret}
\end{figure}
The quark mass in these calculations was about $1.5GeV$.
The discretization errors in the correlation function turns
out to be large, e.g the value of $G(\tau=1/(2 T))$ at $1.5T_c$ 
calculated with $a^{-1}=4.9GeV$ and $a^{-1}=6.5GeV$ differs by $16 \%$
while the same difference is the light quark sector is about $1 \%$.
Because of this MEM cannot be applied to extract the spectral function
unless extrapolation to the continuum limit is performed. Nevertheless,
correlation functions themselves can provide some information about 
existence of quarkonium bound states. In Fig. \ref{ret} the ratio of the
meson correlator calculated at $1.2T_c$ ($48^3 \times 12$, $\beta=6.499$) 
to the one calculated at $0.9T_c$ ($48^3 \times 16$, $\beta=6.499$).
Except the last point at $\tau \simeq 0.3 fm$ this ratio stays very close
to unity at $1\sigma$ level. The deviation at $\tau \simeq 0.3 fm$ is simply
due to periodic boundary condition ($\tau \simeq 0.3 fm$ corresponds to
$\tau = 1/(2 T)$ on $48^3 \times 12$ lattice). While this fact does not
necessary implies that the quarkonium spectral function does not change 
as the deconfinement point is crossed, it does imply that the propagation
of heavy quarks in the deconfined phase is far from free propagation
of $Q \bar Q$ pair in agreement with studies performed in Refs. \cite{fingberg,umeda}.

\section{Conclusions}
I have considered quarkonium dissociation in hot strongly
interacting matter below as well as above the deconfinement. In a
confined
medium dissociation of certain quarkonium states occurs due to in-medium
modification of the open charm (beauty) threshold as well as the
quarkonia masses.
In the deconfined medium quarkonium dissociation is due to color
screening.
In this analysis quarkonium masses were extracted from Schr\"odinger's
equation. The singlet potential used in Schr\"odinger's equation 
was extracted from lattice data on the Polyakov loop correlator 
using some additional assumptions. 
For a more accurate determination of the quarkonium suppression
patterns, it would be desirable to carry out direct lattice studies of
the color singlet potential and of its quark mass dependence, which
may become important near the critical temperature. Furthermore, to make
contact with nuclear collision experiments, a more precise determination
of the energy density via lattice simulations is clearly needed, as is
a clarification of the role of a finite baryochemical potential.
For the latter problem, lattice studies are so far very difficult;
nevertheless, a recent new approach \cite{fodor} could make such
studies feasible.

In section 5 the problem of direct lattice determination of quarkonium
properties was discussed. An unbiased determination of these properties
free of lattice artifacts seems to be very difficult and not available
so far even in the quenched approximation. Such an analysis will be
a very important  check of the results obtained from the simple potential
picture discussed in sections 3 and 4.

\bigskip
\noindent {\bf Acknowledgments:}
\noindent
I am grateful to  A. Patk\'os for reading the manuscript and
making  number of valuable comments. I would like to thank
the organizers of ICPAQGP-2001 for their work in carrying
out this very enjoyable meeting and for their hospitality.
The work has been supported by the 
DFG under grant FOR 339/1-2 and by BMFB under grant 06 BI 902.

\end{document}